\documentclass[showpacs,preprint,aps]{revtex4}
\begin{document}
%
%-----------------------------------------------------------------------
\title{
G-quartet biomolecular nanowires
}
%-----------------------------------------------------------------------
\author{
A.\ Calzolari, R.\ Di\ Felice, and E.\ Molinari
}
%-----------------------------------------------------------------------
\affiliation{
%Istituto Nazionale per la Fisica della Materia, 
INFM and Dipartimento di Fisica, Universit\`a di 
Modena e Reggio Emilia, Via Campi 213/A, 41100 Modena, Italy
}
%-----------------------------------------------------------------------
\author{
A.\ Garbesi
}
%-----------------------------------------------------------------------
\affiliation{
CNR ISOF, Area della Ricerca, Via P. Gobetti 101, 40129 Bologna, Italy
}
%-----------------------------------------------------------------------
\begin{abstract}
We present a first-principle investigation of quadruple helix
nanowires,  consisting of 
stacked planar hydrogen-bonded 
guanine tetramers.
Our results show that long wires form and are 
stable in potassium-rich conditions. 
We present their electronic bandstructure and discuss the interpretation 
in terms of effective wide-bandgap semiconductors. 
The microscopic structural and electronic properties 
of the guanine quadruple helices make them suitable 
candidates for molecular nanoelectronics.  
\end{abstract}
%-----------------------------------------------------------------------
\maketitle
Guanine (G) supramolecular systems receive great 
attention because of their peculiar 
structural and electronic properties: 
G molecules have the capability of forming a variety of self-assembled 
structures (helices, ribbons, tubes), both in solution 
and in the solid state \cite{garbesi,acta}, 
and G-rich DNA sequences exhibit efficient long-range electron 
transfer properties \cite{jortner}. 
Among the different G aggregates, in this letter we focus 
on tubular sequences of G-quartets \cite{garbesi}, 
whose formation is driven by the presence of alkali cations:
such tubes are often labeled as 
G-{\sl quadruplexes} or G-{\sl wires} \cite{afm}. 
The building block of a tube is the tetrameric 
G-quartet (G4) \cite{garbesi}, which is a planar aggregate of four 
hydrogen-bonded guanines arranged in a square-like configuration, 
as shown in Figure 1a. 
Each G4 contains four carbonyl oxygen atoms pointing 
towards the center of the square. By stacking on top of each other, 
these tetramers form a tube-like column with a central cavity, 
that easily accomodates monovalent cations. 
X-ray and Nuclear Magnetic Resonance (NMR) data 
of both hydrophylic and lipophylic G-{\sl quadruplexes}
reveal that 
one cation is symmetrically 
located between any two G4 planes (Figure 1b), and coordinated with eight 
nearest-neighbor O atoms \cite{luisi,gottarelli2}. 
G4 stacks exhibit a surprisingly 
high stability in different chemical conditions 
\cite{garbesi,luisi,gottarelli2}.

Besides the biochemical aspects \cite{garbesi}, 
G-rich systems are attracting 
increasing interest in applied physics
as functional elements 
in molecular electronics \cite{joachim}.
For example, guanine aggregates have been used to 
devise new conducting nanoscale materials,  whose electronic features 
are typical of wide-gap semiconductors
\cite{danny,rinaldi}.
Whereas recognition is a fingerprint of all the nucleic acid bases, 
guanine plays a key role  to enhance charge migration \cite{jortner,schuster},
because of its low ionization potential, and  is 
suitable for electronic applications. 

In this letter, we present a first-principle 
investigation of the structural and electronic properties of 
extended G-{\sl wires}, 
free-standing or intercalated by K$^+$ ions. 
Our results show that in K-rich conditions G4 helices 
are thermodinamically stable, consistently with 
X-ray and NMR experiments 
\cite{luisi,gottarelli2}. 
Although the single-particle bandstructure 
identifies the localized character of the electron wavefunctions, the 
calculated density of states (DOS) is consistent 
with a model of a semiconducting nanowire, which 
hosts extended channels for electron/hole motion. 
The peaks in the effective band-like DOS result from 
the energy spreading 
of the single-molecule energy levels. 
To our knowledge, this is the first 
theoretical prediction of the electronic properties of 
the extended tubular G-{\sl wires}. 

We performed first-principle total-energy calculations based on a
Car-Parrinello-like scheme \cite{fhi}, with all the atoms free 
to relax towards the minimum energy configuration.
The electronic structure is described in the frame of 
plane-wave pseudopotential DFT \cite{dft}, using the
BLYP \cite{blyp} 
exchange-correlation functional:
this scheme gives a correct description of H-bonding 
in molecular systems \cite{water}. 
The accuracy of the computational technique 
was tested on isolated G molecules and on self-assembled G-ribbons \cite{nastri}. 
The {\sl quadruplexes} are simulated by the use of periodically repeated 
supercells (24.3$\times$24.3$\times$10.1 \AA$^3$), 
containing three stacked  C$_4$-symmetric G4 tetramers (Figure 1b), 
separated by 3.4 $\AA$ and
rotated by 30$^{\circ}$ (Figure 1c) 
along the stacking direction. 
The starting atomic configuration in 
our simulations was that resulting from the X-ray analysis 
of the G-{\sl quadruplex} d(TG$_4$T)
\cite{luisi}. 
Neighboring  replicas of the basic structural unit are laterally 
separated by vacuum, while a continuous wire is formed with its axis 
perpendicular to the tetrameric planes. 
Theoretical studies 
showed that the external backbone in helical 
nucleotide structures 
is not involved in transport phenomena \cite{ladik}.
Thus, we believe that our calculations for the G4 stacks 
are well representative of principal electronic features of real systems.

We studied two different G4 columns: 
the tube containing K$^+$ ions along the axis 
in the inner cavity is labeled 3G4/K$^+$ \cite{nota},
the empty tube is labeled 3G4.
The formation energy $E^{form}$ 
of each system, relative to that of isolated G molecules, as a 
function of the Fermi level ($E_F$) and of the chemical 
potential of K ($\mu^{K}$), may be calculated after  
a thermodynamic approach successful for 
the study of defects in semiconductors \cite{vandewalle}, as
$$\Delta E^{form}=\Delta E^{tot}-\Delta n^K\mu^K-\Delta n^eE_F.$$
E$^{tot}$ is the calcutated total energy, 
n$^{K}$  is the number  of K atoms, 
n$^{e}$ is the number of excess electrons with respect to charge neutrality: 
for 3G4 (3G4/K$^+$) $\Delta n^K=\Delta n^e=0$ 
($\Delta n^K=+3$, $\Delta n^e=-3$). 
In the formula, $\mu^{K}$ and E$_F$ 
depend on the 
conditions under which the material is formed 
and must be determined on the basis of physical constraints. 
To obtain Figure 1d, 
we assumed $\mu^{K}=\mu^{K}_{bulk}$, equivalent to fixing 
K-rich conditions. 
Under this assumption, 
the relative formation energy is a linear function of the Fermi level, and 
its slope is determined by the charge state $\Delta n^e$.
Figure 1d shows the relative formation energy of the 
isolated G molecules, the 3G4 and the 3G4/K$^+$ wires, 
as a function of $E_F$: for a given value of $E_F$, 
the lowest energy curve identifies the most favorable phase. 
The plot reveals that the G columns are energetically favorable 
with respect to isolated guanine molecules, due to the stability effect 
of the several hydrogen bonds. 
The 3G4/K$^+$ structure is more stable than the 3G4 structure 
for $E_F$ less than 1 eV: the energy gain 
is 1.8 eV/G4 for $E_F$=0. 
As we discuss in the following paragraph, 
the G-{\sl wires} exhibit an energy bandgap 
between the HOMO and the LUMO, and the condition 
$E_F$=0 corresponds to pinning the Fermi level at the top of the 
occupied energy levels. 
This condition is consistent with the abundance of metal cations, 
that may act as electron acceptors. 
Under these circumstances, 
the high stability of the K$^+$-containing columnar structures
is in agreement with the experimental evidence that the 
{\sl quadruplexes} form only in the presence of metal cations 
\cite{garbesi,luisi,gottarelli2}. 
We find  that the empty 3G4 tube 
is unfavorable with repect to isolated G4 tetramers by 1.4 eV/G4, and
the inclusion of K$^+$ ions 
stabilizes the columnar phase also with respect to the free tetramers.
We attribute this effect to 
electrostatic K$^+$-O$^-$ interactions that counterbalance the strong K$^+$-K$^+$
repulsion \cite{stability}. 
A K$^+$ concentration smaller that 1/G4 would break this equilibrium 
and would not be sufficient to induce the formation of the G-{\sl wires}. 
Our results agree with cluster
quantum-chemistry calculations \cite{gu}.

The electronic bandstructure of the 3G4/K$^+$ tube 
is shown in the right panel of Figure 2, while the left panel 
illustrates the DOS of both the K$^+$-filled (bottom) 
and the empty (top) tubes. 
For other G stacks \cite{nastri}, $\pi$-$\pi$ 
coupling may give rise to delocalized Bloch-type orbitals, 
whose band dispersion depends on the relative rotation angle 
between nucleobases in adjacent planes. 
In the guanine {\sl quadruplexes}, 
the C$_4$ symmetry 
increases the spatial $\pi$-$\pi$ overlap with 
respect to a segment of G-rich B-DNA, and we might 
expect an enhancement of the band-like behavior. 
Instead, our results reveal a different mechanism. 
For both filled and empty G4 columns, we find 
that the inter-plane $\pi$ superposition is not 
sufficient to induce the formation of delocalized orbitals 
along the axis of the {\sl quadruplex}: The right
panel in Figure 2 shows that the bands of the 3G4/K$^+$ 
column are flat, 
typical of supramolecular systems in which the orbitals 
remain localized at the molecules.
The bandstructure identifies the presence of band-multiplets, each
constituted of  12 energy levels.
The  12 electron orbitals associated to a multiplet have identical character 
and localize on the 12 guanines 
in the unit cell.
The energy levels in a multiplet are separated by an average energy difference of 0.02 eV,
smaller than the room-temperature energy k$_B$T: 
therefore, the coupling with the thermal bath 
allows the G-localized neighboring orbitals of a multiplet to interact, 
producing an effective delocalized orbital.
The resulting DOS (Fig. 2), obtained with a gaussian broadering to the calculated one-electron
energy eigenvalues,  shows filled and empty effective bands, separated by a
gap of 3.3 eV and 3.5 eV for the 3G4 and the 3G4/K$^+$
tubes, respectively. 
Each of the five peaks labeled 
with letters in Figure 2 takes origin from one 
G orbital: For instance,  peak {\bf a} is the convolution of 
the $\pi$-like HOMO's of the 12 unit guanines. The amplitude of this peak is 
0.3 eV for both wires. The only meaningful difference that 
we note between 3G4 and 3G4/K$^+$ is a downward shift of 
the $\sigma$-like peak {\bf b} 
in 3G4/K$^+$. By a comparative analysis of a similar shift 
for $\sigma$-like orbitals of the G-ribbons \cite{nastri}, resulting upon the 
introduction of electrostatic interactions,
we attribute the downward shift 
of the $\sigma$-like peak {\bf b} in the 3G4/K$^+$ tubes
to the electrostatic attractive interaction induced by the inclusion 
of the K$^+$ ions.

We show in Figure 3 countour plots of the HOMO convolution 
(peak {\bf a} of the DOS) of 3G4/K$^+$ in a plane containing 
(top) and perpendicular to (bottom) the axis of the quadruple helix. 
The side-view contour plot shows the formation of 
an extended orbital in the outer side of the nucleobase stack, 
due to C-N bonds, that constitutes a channel for electron 
mobility along the helical axis. 
The contour plots also reveal that the K species 
is fully ionized, and therefore the inner cavity of the 
column is free of charge. Thus, conductivity in the 
G-{\sl wires} may not be attributed to metallic bonds between the K atoms, 
neither to their ionic motion, because of the high stability of 
the structure. 

By means of first-principle calculations, we have demonstrated 
the stability of columnar structures based on guanine stacks, 
in the abundance of K$^+$ ions. The $\pi$-$\pi$ coupling 
does not induce extended states and dispersive energy bands.
However, the energy separation between neighboring localized states is 
so small that coupling can be easily induced, {\sl e.g.}, by room-temperature 
thermal hopping.
Therefore,  3G4/K$^+$ 
tubes under suitable conditions,
are expected to exhibit an effective behavior of 
wide-bandgap semiconductors. This feature, along with the 
possibility of forming extended stacked wires at the 
nanoscale length, makes them appealing for the development of 
biomolecular electronics.

This work was supported  by INFM through
{\sl Progetto Calcolo Parallelo}, and PRA-SINPROT.
%%%%%%%%%%%%%%%%%%%%%%%%%%%%%%%%%%%%%%%%%%%%%%%%%%%%%%%%%%%%%%%%%%%%%%

%-----------------------------------------------------------------------
%
\begin{figure}
\caption{(a) Schematic representation of the planar G4; (b) G-{\sl wire}:
side view of the three repeated tetramers in presence of K$^+$ ions; 
(c) schematic top view of the stacked G4's in the
helical wire; (d) relative formation energy  vs. Fermi level, $\Delta$E$^{form}$=0 corresponds to the chemical potential of the single guanine, E$_F$=0
marks the top of the valence band.}
\end{figure} 
\begin{figure}
\caption{The left panel illustrates the DOS of the  periodic G4 stacks with (bottom) and without (top) K$^+$ ions.
 The right panel shows the almost dispersionless energy levels of the 3G4/K$^+$ system. 
$\Gamma$ (A) is the center (edge) of the 1D
 Brillouin Zone  in the stacking direction.
Peaks in the DOS are labeled to identify their 
common microscopic origin.}
\end{figure} 
\begin{figure}
\caption{Contour plots of the symmetrized charge density of the HOMO in two different planes containing (top) and cutting (bottom) the axis of the G4 stack.
Top: the vertical direction coincides with the axis of the stack, the horizonal direction is prependicular to such axis and goes through the atoms labeled in the figure. Bottom: the fixed-density contours are
shown in the plane containing the middle G4 assembly of Fig. 1b.}
\end{figure} 
%-----------------------------------------------------------------------

\begin{thebibliography}{00}
\bibitem{nota} 
The {\sl quadruplexes} 
characterized by X-ray contained Na$^+$ ions rather than K$^+$ ions: 
we assumed the same starting positions for the K$^+$ ions.
\bibitem{stability} 
Along the axis of a G{\sl-quadruplex}, the potassium ions
are separated by a distance of 3.4 \AA, smaller than the
interatomic distance in the bulk $fcc$ metal.
\end{thebibliography}
\end{document}